# Large-scale identification of coevolution signals across homo-oligomeric protein interfaces by Direct Coupling Analysis

Guido Uguzzoni[1]*, Shalini John Lovis[2]*, Francesco Oteri[1], Alexander Schug[2]+, Hendrik Szurmant[3]+, Martin Weigt[1]+

**Affiliations**:
1. Sorbonne Universités, UPMC Université Paris 06, CNRS, Biologie Computationnelle et Quantitative - Institut de Biologie Paris Seine, 75005 Paris, France
2. Steinbuch Centre for Computing, Karlsruhe Institute of Technology, Hermann-von-Helmholtz-Platz 1, 76344 Eggenstein-Leopoldshafen, Germany
3. Department of Basic Medical Sciences, College of Osteopathic Medicine of the Pacific, Western University of Health Sciences, Pomona CA 91766, USA

* joined first authorship
+ to whom correspondence might be addressed

**Corresponding authors**:
Alexander Schug: Hermann-von-Helmholtz-Platz 1, 76344 Eggenstein-Leopoldshafen, Germany, tel. +49 721 608-26303, email: schug@kit.edu

Hendrik Szurmant: 309 E. 2nd Street, Pomona, CA 91766-1854, USA, tel. +1-909-706-3938, email: hszurmant@westernu.edu

Martin Weigt: LCQB UMR7238, 4 place Jussieu 75005 Paris, France, tel. +33 1 44 27 73 68, email: martin.weigt@upmc.fr

**ABSTRACT (250 words or less)**

Proteins have evolved to perform diverse cellular functions, from serving as reaction catalysts to coordinating cellular propagation and development. Frequently, proteins do not exert their full potential as monomers but rather undergo concerted interactions as either homo-oligomers or with other proteins as hetero-oligomers. The experimental study of such protein complexes and interactions has been arduous. Theoretical structure prediction methods are an attractive alternative. Here, we investigate homo-oligomeric interfaces by tracing residue coevolution via the global statistical Direct Coupling Analysis (DCA). DCA can accurately infer spatial adjacencies between residues. These adjacencies can be included as constraints in structure-prediction techniques to predict high-resolution models. By taking advantage of the on-going exponential growth of sequence databases, we go significantly beyond anecdotal cases of a few protein families and apply DCA to a systematic large-scale study of nearly 2000 PFAM protein families with sufficient sequence information and structurally resolved homo-oligomeric interfaces. We find that large interfaces are commonly identified by DCA. We further demonstrate that DCA can differentiate between subfamilies with different binding modes within one large PFAM family. Sequence derived contact information for the subfamilies proves sufficient to assemble accurate structural models of the diverse protein–oligomers. Thus, we provide an approach to investigate oligomerization for arbitrary protein families leading to structural models complementary to often-difficult experimental methods. Combined with ever more abundant sequential data, we anticipate that this study will be instrumental to allow the structural description of many hetero-protein complexes in the future.

**SIGNIFICANCE STATEMENT (<120 words)**

Protein-protein interactions are important to all facets of life, but their experimental and computational characterization is arduous and frequently of uncertain outcome. The current study demonstrates both the power and limitation to study protein interactions by utilizing sophisticated statistical inference technology to derive protein contacts from available sequence databases, more precisely from the coevolution between residues, which are in contact across the interaction interface of two proteins. By studying homo-oligomeric protein interactions the current study expands from anecdotal evidence of the performance of this technology to a systematic evidence of its value across close to 2000 interacting protein families.

**INTRODUCTION**

Life on the molecular level is orchestrated by the interplay of many different biomolecules. A crucial component for the function is its structure, ranging from small monomers to complex homo- or hetero-multimers. The full structural characterization of a biomolecule therefore typically precedes a detailed explanation of their functional mechanism. Yet despite the incredible progress of structural characterization methods, many important biomolecules have not been structurally resolved. An intriguing alternative to often involved experimental measurements of three-dimensional structures is taking advantage of the growing wealth of genetic sequential information via sophisticated statistical methods. Direct-Coupling Analysis (DCA) (1, 2) and related tools (3, 4) develop a global model mimicking evolutionary fitness landscapes of protein families (5, 6) and quantify the coevolution of amino acid residue positions(7-9). In the context of protein structure prediction, these models allow extraction of residue-residue contacts from sequence information alone. Such information has proven useful in the prediction of tertiary protein structures (10-13), conformational transitions (14-16), and similarly RNA structures (17, 18).

To be successful, global statistical models must be trained on sufficiently large alignments of homologous sequences. It has been argued (2) that on the order of 1,000 properly aligned and sufficiently divergent sequences are adequate for accurate model learning. Scanning the latest Pfam database release (19), we observe a fundamental extension of the families amenable to statistical modeling, cf. SI Appendix, Fig. S1: While in the Pfam5.0 release in the year 2,000 only 33 families contained at least 1,000 sequences, this number has grown by a factor of more than 200 to 6,783 in Pfam28.0 (released in 2015), out of included 16,230 families. The median family size of 654 sequences is close to the required sequence number. Statistical sequence modeling will therefore strongly increase in importance over the next years.

Statistical modeling of single protein families is the final aim: proteins rarely work in isolation and monomeric form. Protein-protein interfaces coevolve to conserve the interaction and interaction specificity between proteins. To identify such coevolving protein-protein interfaces has been the original motivation for DCA development (1, 20), and this has recently been extended beyond simple anecdotal cases to dozens of protein pairs (13, 21-25). Much more extensive datasets are needed for a solid statistical assessment of the quality of coevolution-based predictions, which tells us when DCA can reliably predict residue contacts between interacting proteins and how to overcome a possible failure. The main difficulty in going to much larger data sets results stems from the necessity to create joint multiple-sequence alignments (MSA) of interaction partners, where each line contains an interacting pair of proteins. Computationally this becomes a hard problem by itself due to the existence of paralogs (26-28), and most works so far have concentrated on systems where interaction partners have co-localized genes, e.g. in common operons in the bacteria.

The present work addresses the challenge by focusing on *homo-oligomerization* rather than hetero protein-protein interactions thereby increasing the number of available data sets by more than one order of magnitude. The reason is simple: The creation of joint multiple-sequence alignments (MSA) of interaction partners is simplified, since the identical sequence of both interaction partners allows to work with MSA for single proteins. On the contrary, the distinction between intra-protein and inter-protein residue-residue contact predictions from coevolutionary couplings is non-trivial when studying homo-oligomers, and the existence of experimental monomer structures is needed to identify putative inter-protein contacts as directly coevolving pairs with long intra-monomeric distance.

By coupling DCA with *in silico* molecular simulations we propose an analysis approach that can be applied to modeling both homo-oligomeric as well as hetero-oligomeric biological assemblies of protein complexes (Fig. 1). In this workflow, contact-residue pairings between proteins are extracted from sequence alignments of large protein families. Intra-protein contacts are eliminated by pruning those DCA-predicted contacts compatible with known monomeric structures. The remainders of pairings are considered as potential inter-protein contacts and are utilized to drive docking of monomeric structures into biological assemblies. Where individual proteins have not been structurally resolved, these can typically be homology modeled.

The concentration on homo-oligomers allows us to go substantially beyond the current stage of coevolutionary analysis of interacting proteins by analyzing nearly 2,000 systematically selected and sufficiently abundant protein families (i.e. MSA large enough to provide the statistical signal detected by DCA) with known structures (Fig. 2). This number of families makes the use of automated procedures essential. Together with the strict selection criteria of investigated protein families, this approach reduces possible artifacts coming from a subjective selection of systems and/or human data curation. The analysis of more than 750 families with large interfaces show significantly enriched coevolutionary couplings across the interfaces, while small interfaces are rarely detected. However, even for large interfaces the detection of strong coevolutionary couplings and therefore the accuracy of the resulting contact predictions are limited to roughly 50% of cases. To better understand this observation, and to open paths for overcoming this limited performance, we analyze in more detail the specific case of response regulator proteins involved in bacterial two-component signal transduction, where several subfamilies (characterized by different domain architectures) show different homo-dimerization modes. We show that, due to the non-conservation of the dimerization mode in the entire Pfam (19) domain family of response regulators, a straight-forward application of DCA leads to a weakened signal as different dimerization modes are mixed. The restriction of the MSA to proteins of presumably the same dimerization interface, however, strongly improves the strength and accuracy of the coevolutionary signal. We demonstrate on the case of response regulators that by subdividing the alignments based on protein domain architecture sufficient contact information can be extracted from sequence to reliably and accurately model the dimer structures *in silico*, while the coevolutionary signal in the entire protein family is dominated by the common structural characteristics of all subfamilies (mostly tertiary contacts).

Finally, we analyze the 142 upon the nearly 2,000 large Pfam families without an annotated biological assembly. Interestingly, only a small number of these show a strong non-monomeric coevolutionary signal. Individual analysis of these cases does not provide consistent oligomerization signals, suggesting that to date many of the physiologically relevant homo-oligomerization structures with large and well-conserved interfaces in sufficiently populated protein families have already been experimentally described. In addition to this important finding, our procedure is applicable to the determination of hetero-protein complexes and in the detection of alternative homo-oligomerization modes.

**RESULTS**

*Database selection of biological oligomers*
This study aims to assess how readily interaction surfaces of proteins can be identified from large protein sequence alignments, by analyzing coevolutionary signals between residue positions. To accomplish this systematically we focused on homo-oligomer interactions. To this end we scanned the Pfam27.0 database (19) for domains with sufficient sequence diversity to apply DCA, as outlined in the *Methods* section. In addition we required at least one high-resolution protein structure to be deposited in the Protein Data Bank (PDB) (29) in order to distinguish coevolving intra- from inter-protein contacts as further discussed below. Our study thus provides an exhaustive analysis of coevolution across homo-oligomeric interfaces.

The above requirements resulted in a dataset of 1984 protein domain families (Fig. 2). The dataset was further analyzed for the existence of an annotated biological homo-oligomeric assembly. Interestingly, only 146 protein families within the dataset had no annotated biological assembly, suggesting that the large majority of protein families are believed to form dimers or higher order structures of physiological relevance. Protein families with known biological assemblies were further subdivided based on size of the interface covered by the Pfam sequence models, since larger interactions surfaces can be expected to be more readily identified than those with few interaction contacts. The 1838 protein families with annotated biological assemblies were thus further subdivided into 750 families with large and 1088 with small biological assembly interfaces as detailed in *Methods*. The 750 families with large biological interfaces are our main focus of this manuscript but the other families were also analyzed for coevolutionary signals.

*Intra- versus inter-protein contacts in homo-oligomers*
The coevolutionary detection of inter-protein contacts in homo-oligomeric assemblies is, at the same time, easier and more complicated than the prediction of inter-protein contacts in hetero-oligomers, cf. Fig. 1. It is easier, since we need a joined multiple-sequence alignment (MSA) of both interaction partners across the interface, which in the case of homo-oligomers are identical. It is therefore sufficient to have one MSA of a single protein. This, however, implies that the differentiation of intra-protein and inter-protein contacts becomes non-trivial in the case of homo-oligomers. Without any exemplary monomeric structure, this problem remains currently unsolved. When an exemplary structure of the monomer exists – and this is the case treated in this work – we can interpret all those residue pairs as inter-protein contact predictions, which are distant inside the monomer, but show a strong coevolutionary signal. In other words, not realized "false-positive" intra-protein predictions are interpreted as possible inter-protein predictions. As illustrated in supplementary Fig. S1, this may hide part of the protein-protein interface: In particular in symmetric assemblies, residue pairs may be in spatial vicinity both inside the monomer and across the interface of the complex. Actually, while only about 14.5% of all residue pairs are within 8Å inside a single protein, we find about 32.3% of all inter-protein contacts (8Å cutoff) to be also intra-protein contacts (5.4% vs. 14.8% at a more stringent 5Å distance cutoff). This may possibly happen in alternative structures, e.g. by the mechanism of domain swapping (30), where internal contacts of the monomer become interface contacts in the dimer. Disentangling the two is, again, a currently unsolved problem, and their physiological relevance is not always clear.

To shed light on the problem of hiding parts of the interface when concentrating on large internal distances, we typically use more than one distance cutoff for identifying intra-protein contacts. As a general tendency, we see that residue-pairs with a strong coevolutionary signal, but at large distance in the monomer (e.g. minimal atom-distance above 12Å) have a very large probability of being inter-protein contacts, cf. Fig. 3. Pairs at lower distances, but not directly in contact (e.g. between 8Å and 12Å) inside the monomer, might be accommodated as internal contacts by minor conformational changes or different residue choices, and have thus a lower probability of being actually inter-protein contacts.

*Strongly coevolving residue pairings are typically internal or inter-protein contacts*
To assess the accuracy of contact prediction in homo-oligomers, we selected the database of 750 protein-domain families which feature (a) sufficient sequence space to enable statistical analysis, (b) have at least one high-resolution structure deposited in the protein data bank annotated as biological homo-oligomer and (c) whose homo-oligomer interfaces are of sufficient size, cf. *Methods*.

When analyzing these 750 protein families, the first expected result is that most strongly coupled residue pairings represent intra-protein contacts defining the fold of the domain family (Fig. 3A). When only considering internal contacts the positive predictive value (PPV) defined as true positive contact predictions over all predictions (TP/(TP+FP)) for all 750 families shows 90.2% accuracy for the top 10 highest scoring contact pairings. When also considering inter-protein contacts the PPV is further increased to 95.2%. In other words, on average 95.2% of the top 10 highest scoring residue pairs across 750 protein families are either internal or inter-protein contacts. When analyzing residue pairings according to their coevolutionary score $F_{APC}$ (see methods and ref. (31)) we find that pairings with a score over 0.3 have a PPV of about 90% (Fig. 3C). In other words, a score $F_{APC}>0.3$ provides an excellent prediction that two residues are in contact, thus providing a natural prediction cut-off across diverse protein families.

*Enrichment of coevolutionary couplings across homo-oligomeric interfaces*
The accuracy of our inter-protein contact prediction was further analyzed by two main measures. First, for each of the selected families, we extract the *n* pairs of strongest coevolutionary signal, which are not in contact inside the monomer (cf. *Methods)*, and determine the fraction of these predictions being inter-protein contacts in biological homo-oligomeric assemblies. Fig. 3B shows this quantity as a function of the number $n$ of inter-protein contact predictions. We observe a clear enrichment in true inter-protein contacts; more than 40% of the first inter-protein predictions are in contact in at least one of the considered structures, as compared to 2.9% of all residue pairs. However, despite this strong enrichment, we observe also that up to 60% of the first predictions are actually false positives, i.e. they are neither in contact in the monomer nor in the oligomer. To better understand this finding, we focused on the PPV according to the coevolutionary score $F_{APC}$. Sorting all internal non-contacts according to their scores, we find again that values above $F_{APC}>0.3$ have a very high probability (>60%) of being in contact across the interaction interface, in particular if they are distant inside the monomer, (Fig. 3D). The PPV even grows to ca. 80% for $F_{APC}>0.5$. Not all protein families show, however, such strong coevolutionary coupling: While a vast majority of families (714 out of 750) show at least one residue pair with internal distance above 8Å and coevolutionary score $F_{APC}>0.2$, this number drops rapidly when going to larger scores: 483 families have at least one coupling with $F_{APC}>0.3$, 190 with $F_{APC}>0.5$. As stated before, not all of these pairs are necessarily quaternary contacts. Asking, e.g., for at least 5 (resp. 10) internal non-contacts with $F_{APC}>0.3$, we find only 236 (resp. 111) families, for $F_{APC}>0.5$ this number reduces to 55 (resp. 16). The full dependence of the cutoff of the DCA-score is given in SI Appendix, Fig. S2.

*The size of the homo-oligomeric interface significantly influences the availability of highly correlated interface positions*

The high probability of being an inter-protein contact if a large coevolutionary coupling was observed together with elevated intra-protein distance, is very interesting: It underlines the observation that most direct couplings unveiled by DCA are actually based on physical contact, and no other mechanisms (like, e.g., allosteric coupling), which in consequence should be mediated by spatially connected networks of intermediate residues. However, a substantial number of protein families do not show any strong homo-oligomeric DCA signal. To better understand this fact, we have partitioned the data set according to the size of the interface. In Fig. 4A, protein families are divided into three equally large groups containing the smallest, the largest, and intermediate interface sizes, and histograms of the PPV (positive predictive value) for the first 20 homo-oligomeric contact predictions are displayed. We see that small interfaces have by far the largest probability of remaining undetected, while interfaces of high PPV are dominated by large interfaces. We conclude that the larger the interface the more readily it might exhibit strong coevolutionary signals and that smaller interfaces do not necessarily have the same requirement for a significant coevolutionary signal.

*The conservation of the homo-oligomeric interface significantly influences the availability of highly correlated interface positions*

In Fig. 4B, we have analyzed the data set according to the interface conservation, cf. SI Appendix, Sec. SI-3, for a precise description, how this has been obtained. As a general idea, an interface is considered to be conserved if, over different available PDB structures, interfaces tend to strongly overlap. On the contrary, it is considered to be non-conserved, if entirely different homo-oligomeric interfaces are found in the PDB. Dividing again the dataset into three classes, we see that more conserved interfaces have larger PPV than variable ones. This could be expected: A pair being in contact in one specific interface may coevolve in all proteins showing this interface, and evolve independently in all proteins showing alternative or no interface at all. Therefore, the coevolutionary signal for each individual interface becomes diluted if we consider the full Pfam MSA, and thus less easily detectable by DCA. We hypothesize that by focusing on a sub-alignment possessing a common oligomerization interface, the coevolutionary signal might become more evident – as long as the sub-alignment contains sufficient sequences. In fact, the issue of differing interaction modes might be a primary reason why DCA fails to predict inter-protein contacts in many protein families that fulfill requirements of sequence availability and interface size since Pfam protein families commonly include proteins with similar folds but diverse function and activity.

*Response regulator dimers: Non-conserved dimerization modes and subfamily specific coevolutionary signals*

To test the hypothesis on differing interaction modes we focused on one particular protein domain family with a known diverse set of protein-protein interfaces, the bacterial response regulators (RR) defined by Pfam ID PF00072. It is an extremely abundant protein family used predominantly in bacterial signal transduction. RR domains are subject to phosphorylation by an associated histidine kinase. In response to phosphorylation they alter their activity most commonly by acting as transcription factors. Diverse DNA-binding domains are utilized by these RR proteins and accordingly, different subclasses oligomerize in different ways, cf. Panels C,H and I of Fig. 5. Some subclasses of RR are of single domain architecture and thus do not have additional output domains but rather act directly by forming hetero-protein interactions, e.g. those that mediate bacterial chemotaxis as reviewed in (32).

The different DNA-binding domains found in transcription-factor RR belong to different Pfam families. Out of the 151,337 RR sequences listed in Pfam, 45431 have a Trans_Reg_C transcription-factor domain, 20,829 a GerE domain, and 6,269 a LytTR domain. Other architectures (RR/Sigma54_activat/HTH_8,RR/HTH_18) are less frequent, besides the single-domain architecture, which covers 10,673 protein sequences. Whereas the activated dimer structure of the Trans_reg_c associated RR has long been known (33, 34), dimer structures of LytTR and GerE associated RR have only more recently been determined (35, 36). All three classes of RR show significantly different dimerization modes (Fig 5., right panels).

In Fig. 5, we show contact maps for all three classes, overlaid with DCA prediction, both for intra-protein (upper diagonal) and inter-protein contacts (lower diagonal). The left panels (A,D,G) show the predictions from the full Pfam MSA. The inter-protein prediction shows a mixture of contacts from different architecture, with a preference for the dominant Trans_Reg_C class. Consequently, for each individual architecture many false positives are found. The picture changes substantially, if we consider architecture-specific sub-alignments (Panels B,E,H). The most evident change is the elimination of many FP predictions resulting from contacts in other architectures. The best TP signal is observed in the largest architectural sub-family, the Trans_reg_C class (1, 2, 20). Substantial improvements are also obtained in the GerE-containing architecture. Note that for the third, smallest sub-MSA of the RR/LytTR architecture, a smaller improvement in terms of correctly predicted contacts is discovered. However, it has to be noted that the majority of native inter-protein contacts in this case is also in close spatial vicinity inside each protein monomer (blue in Fig. 5), and thus hidden from our analysis as discussed before.

In summary, by analyzing the RR in detail, we find that by subdividing a specific family of proteins according to their domain architecture we can strongly improve the predictive value of DCA. It can be expected that this is true for many protein families where DCA currently fails to detect an oligomerization signal. In addition, the data demonstrate for the RR protein family, how one particular protein fold can evolve to accommodate various different interaction modes according to need, which in this case is connecting the homo-dimerization mode with the various different DNA-binding domains.

*Coevolutionary analysis on RR subalignments provides sufficient information for accurate structural prediction of protein-complex structures.*
We next ask whether the contact information forthcoming from coevolutionary analysis of the three architecture-specific sub-MSA is sufficient to dock monomeric structures and to determine the different biological assemblies. To this end we utilize the above determined cut-off of $F_{APC}>0.3$; all residue pairs with $F_{APC}>0.3$ and intra-protein distance $d_{intra}>8Å$ are listed in Table 1. We consider them putative contacts (whether realized or not in the known dimer structures), they are utilized in HADDOCK (37) docking studies to determine whether accurate structural models of homo-dimers can be inferred from sequence derived contact information, cf. *Methods* for details.

HADDOCK does not use directly the DCA-paired residues as contacts, but declares all residues in the pairings as potential interface residues. Therefore, docking results in several clusters of structures for each protein subfamily. DCA pairings can be used to rank clusters by the number of actually realized putative contacts, the first-ranking is considered as the likely physiologically relevant dimer. When comparing these dimer models with the experimental structures we observe excellent agreement for all three subclasses of RR proteins (Fig. 6). The mean deviation for each structural dimer cluster in respect to the dimer structures was an impressive 1.1 Å for Transreg_c, 1.2 Å for LytTR and 1.1 Å for GerE subfamilies, respectively. For each of these structures, several of the DCA-predicted residue pairs are realized as inter-protein contacts. Others are not, they all constitute residue pairings with an intra-protein distances 8Å < $d_{intra}$ <12Å and thus likely represent intra- rather than inter-protein contacts that might be realized in other members of the protein family, or by conformational changes between inactive RR monomers and active RR dimers.

We note that GerE has surprisingly two annotated biological dimers in the PDB, one author assigned and one PISA-software assigned. We reconstitute the latter, suggesting that this is the physiologically relevant dimer for GerE type RR, whereas the author-assigned dimer is likely irrelevant.

We realize that the astonishing accuracy of the predicted dimer models in respect to experimental structures represents a best-case scenario, since the monomers used in the docking approach actually derived from physiological dimers. In the SI, we also explore for the Trans_Reg_C subfamily what could be considered a worst-case scenario, where the RR monomer structure was first homology modeled based on the structure of a monomeric homolog, and then used for docking studies. This resulted in slightly reduced but still impressive accuracy (cf. SI Appendix, Sec. SI-4, Fig. S4-S8, Tab. S1, S2). As detailed in the methods and the supplement, we also compare the accuracy of HADDOCK docking with that obtained by MAGMA (20, 38) using $_E$SBMTools (39). With MAGMA we realized the specific DCA-pairings (cf. SI Appendix, Fig. S6, Tab. S1, S2). With HADDOCK, in a first step, all residues in the DCA-pairings are considered as potential interface residues ("non-specific pairings") and then the model with the highest number of DCA-pairings is selected as the best prediction. Little difference in model accuracy was obtained with these substantially differing docking approaches demonstrating that the accuracy of our approach is compatible with several different protein docking techniques.

In summary, these data demonstrate that sequence derived contact information for subfamilies of protein folds can be utilized successfully to predict alternative physiological oligomerization modes for one large but diverse protein family. We anticipate that there are many other protein families where some but not all relevant oligomerization modes are currently known, and where this approach would be of significant utility. In addition, as evident for the example of GerE, a DCA approach represents a complement to PISA interaction software to determine the relevance of a specific oligomerization mode.

*Coevolutionary signal in families without known homo-oligomeric crystal structures*
In our selection of homo-oligomeric structures, we have seen that almost all those residues pairs are actually homo-oligomeric inter-protein contacts, which show a strong DCA signal while being distant in the protein's tertiary structure. Strong DCA signals appear to be, with high probability, related to residue contacts.

At this point an interesting question appears: How many strong signals are there in protein families without known homo-oligomeric structure? To this end we have analyzed the 1234 protein families, which satisfy our selection criteria for sequence abundance and availability of high-resolution monomer crystal structure, without having a sufficiently large homo-oligomeric interface in the PDB. 786 families have no residue pairings with $F_{APC}>0.3$ at intra-protein distances above 12Å, and 235 families have only one such large DCA signal. Out of the remaining 213 families, 189 are actually listed as homo-oligomeric biological unit in the PDB, but have been excluded by or selection criterion on interface quality and size, only 24 have no known homo-oligomeric biological unit. This means that – over all sufficiently large Pfam families with available high-resolution structures – there are very few directly coevolving pairs, which are not explainable via known intra- or inter-protein contacts. In the *Supplement*, we provide a list of all 448 families with at least one distant high-DCA scoring residue pair, together with these pairs, for further exploration.

**DISCUSSION**

The rapidly expanding protein-sequence databases provide us with the necessary raw material to explore the interplay between the evolutionary sequence variability in homologous protein families, and the structural and functional aspects of proteins, which are conserved across species. Coevolutionary modeling approaches – in particular global modeling approaches as represented by the Direct-Coupling Analysis and related methods – have played an important role in detecting residue-residue contacts from directly statistically coupled residues. In the context of tertiary-structure prediction of proteins, this has been tested across hundreds of protein families, and in turn has led to a large number of in-silico tertiary-structure models (10-13) for a large variety of protein families, each one containing thousands of proteins but not having a single experimentally solved representative structure.

The success remains more limited in the case of protein-protein interactions, even if DCA was originally conceived for this case. The largest-scale studies remain so far limited to about 100 protein families (21). The reason is simple; coevolutionary analysis requires large multiple-sequence alignments as inputs, with each line containing a pair of interacting proteins from the two interacting families under study. Due to the abundance of paralogs in a large fraction of protein families, the generation of such alignments is a hard task, and studies typically have been restricted to solvable cases where interacting proteins are co-localized along the genomes, e.g., in operons in the case of bacteria.

In this paper we extend the analysis by about one order of magnitude concentrating on *homo-oligomeric* interactions. Since both interaction partners have identical sequences, the analysis can be done on a single-protein alignment. However, as a potential problem we have identified that, even in the case of the presence of monomeric protein structures, the distinction between intra- and inter-protein contacts is frequently non-trivial. Many residue pairs are in contact both inside the protein and in between the proteins: Large coevolutionary scores would be coherent with the monomeric structure, and no existing method might identify them as inter-protein contacts. They are *a priori* hidden for our method, but in principle they might be used *a posteriori* as a support for a docked oligomer model.

Despite this complication, we find that, out of the close to 2,000 protein families having sufficient sequence numbers and experimentally determined high-resolution crystal structures, only about 150 families have no assigned oligomeric biological unit, and about 750 have a sufficiently large interface to be potential targets of coevolutionary analysis. This large data set allows us, for the first time, to carefully assess the strength *and* the limitations in applying methods like DCA to protein-protein interactions ensuring reproducibility of results and avoiding bias.

First, we find that the majority of "false-positive" predictions of intra-protein contacts, i.e. residue pairs of high DCA score but large distance in the protein monomer, are actually inter-protein contacts involved in homo-oligomerization. Remarkably, we find that almost all residue pairs with coevolutionary scores $F_{APC}$ above 0.3 are either intra- or inter-protein contacts (or both), demonstrating that the strongest couplings detected by DCA are contacts, and not related to any other structural or functional aspects.

We also observe that not all families have such large coevolutionary scores corresponding to distant residue pairs in the monomer structure. Half of the considered 750 protein families with sufficiently large interfaces have few or weak oligomerization signals, and DCA fails to predict the oligomerization mode right away. However, we also find that, the larger the interface is, the higher is also the probability of being detectable by DCA, while smaller interfaces are frequently missed. This raises immediately the question if methods like PconsC2 (40), which uses machine-learning ideas to combine DCA scores with contact patterns from existing protein structures (thereby discriminating locally coherent but potentially weak coevolutionary signals from strong but incoherent noise), can be adapted from intra-protein to inter-protein contact prediction.

We were able to identify a second limiting factor for the capacity of DCA to detect inter-protein contacts. In many protein families the homo-oligomeric complex structure is not conserved, even if the monomeric structure is almost unchanged. In these cases, the coevolutionary signal for each oligomeric structure is present only in part of the large MSA; it becomes weaker if the full MSA is analyzed. This general observation has motivated us to study, in detail, the case of bacterial response regulators. A focus was on those acting as dimeric transcription factor when activated. In this case, different DNA-binding domains correspond to different dimerization interfaces even in the common RR domain. When restricting the MSA to a domain-architecture specific subMSA, specific oligomerization signals emerged. These signals were, for each of the three dominant domain architectures (Trans_reg_C, GerE and LytTR class), sufficiently precise to guide docking procedures to extremely high precision. When analyzing the full MSA, oligomeric signals faded out and mixtures of the different oligomeric interfaces where found. We see that understanding the origins of the limited accuracy of DCA in predicting inter-protein contacts in the full Pfam MSA actually opens up a strategy for a finer-scale analysis based on subfamilies. In future work, it will be interesting to explore in more detail similar cases, and the emergence of subfamily-specific signals when reducing the alignment depth.

Finally, we analyze the 142 families without an annotated biological assembly. Interestingly, only a small number of these expresses a strong coevolutionary signal. Individual analyses of these cases do not provide consistent oligomerization signals. Therefore not a single example of known tertiary but unknown quaternary structure is forthcoming from this study that would allow for *in silico* structure prediction. This suggests that to date many of the large and widely conserved homo-oligomerization interfaces – those where we would expect a clear coevolutionary signal – have already been experimentally described.

Experimental studies of protein structures are by their nature slow throughput and usually involve a single or at best a handful of examples of a specific family of proteins. The scientific literature is abundant with generalizations of protein interactions based on a single structure as representative for an entire protein family. One strength of DCA is that it collects all examples of a given protein family in one MSA, and forthcoming interaction data can thus be considered to be truly representative for most of the proteins of a given family. This fact presents another unique utility of DCA in aiding the researcher to determine whether broad generalizations are indeed possible based on an individual protein structure and its interactions, or if appropriate subfamilies have to be extracted.

The case of GerE suggests the potential utility of DCA in identifying physiologically relevant interfaces in protein crystals structures. Out of the two deposited alternative biological GerE assemblies, only one shows strong coevolutionary couplings across the interface, suggesting a physiological relevance resulting in selective pressure in evolution. The global analysis presented here does not take this possibility into account systematically, a residue pair is considered a true positive if found in contact in at least one assembly (i.e. in the union of all contact maps). However, we can refine the analysis to systematically search for distinct interfaces, and to diversified DCA results for the different interfaces. We have done this for each PDB independently, and kept cases where *(i)* at least 5 pairs with high non-tertiary coevolutionary signal ($F_{APC}>0.3$, $d_{intra}>12Å$) exist, *(ii)* one interface has more than 50% true contacts within these predictions and *(ii)* a distinct interface has less than 10% true contacts. We find 27 cases, mostly with distinct interfaces inside a higher-order oligomeric assembly. A simply understandable case is given by PDB structure 1EA4 (41), cf. SI Appendix, Fig. S9: the structure consists of several DNA-bound dimers. The internal interface in each dimer is well predicted by DCA (6 pairs with $F_{APC}>0.3$, all in contact), the interface between dimers shows no signal and is generated probably only by the DNA functioning as a joint scaffold. While this last result would have a probability of 0.81 to appear in 6 randomly selected residue pairs, the 6 inter-protein contacts are highly significant ($p\sim10^{-5}$).

A full analysis of different homo-oligomer interfaces should compare also different PDB structures for the same protein family. However, caution is in place: as shown again in the case of RR, only the major subfamily (Trans_reg_C) has some small but detectable dimerization signal on the level of the full Pfam alignment, while interfaces signal corresponding to minor subfamilies become invisible in the full MSA. A careful study of individual cases would be needed to be able to distinguish cases where coevolutionary signals are completely absent for a crystallographic interface; or they are present only in a limited subfamily.

Notably, by focusing on a large data set we defined a score cut-off $F_{APC}>0.3$ applicable across all protein families as an excellent determinant whether residue pairs can be expected to make contact in the physiological structure. We expect this cut-off to be of significant value when expanding the current efforts to the identification of novel hetero-protein-protein interactions and this is one of our future goals. As mentioned in the text earlier, due to the existence of amplified protein folds and paralogs, identifying the correct interaction partners for accurate generation of joint MSA is not trivial and needs to be solved in order to apply this technology to hetero-interactions on a global level.

**METHODS**

*Selection of protein families and oligomeric structures*
To perform a large-scale analysis of coevolution in homo-oligomers, we have selected an exhaustive database of protein families from Pfam 27.0 according to the two following criteria:

- Families are required to have, at 80% sequence identity, an effective sequence number of at least 500 (the definition of the effective sequence number follows Morcos et al. (2)), to guarantee sufficient statistics to detect coevolutionary signals by DCA.
- At least one high-resolution (<3Å) structure with homo-oligomeric contacts in the biological unit is present in the PDB.

This last step requires some more precise description. Many protein families have more than one homo-oligomeric interface classified as biological assembly in the PDB. These assemblies may be quite diversified, and a DCA prediction being actually in contact in at least one PDB can be considered as a true-positive prediction. To this end, we have, for each Pfam family

- Collected all PDB with a biological assembly that contains homo-oligomers between chains matching the Pfam family.
- For each PDB, domain repeats in the same chain and different assemblies are taken into account. At this point, a homo-oligomeric assembly is uniquely characterized by the following list: (Pfam ID, PDB ID, chain 1, chain 2, chain 1 domain number, chain 2 domain number, biological assembly number).
- We create a mapping between the Pfam HMM and each concerned chain, which allows us, for each pair of matched columns in the Pfam HMM, to calculate intra- and inter-protein residue-residue distances. Distances are measured as minimal distances between heavy atoms.
- PDBs of low sequence resp. interface coverage by the Pfam MSA (<30% HMM positions matched, <15 interface residues (contact distance <8Å)) are excluded from further analysis.
- Distances between two positions (columns) in the Pfam MSA are now defined as the *minimum* over all matched native distances in all retained PDB files, for intra-protein and inter-protein distances respectively.
- A last step of filtering removes remaining small interfaces of very low contact density $d_{inter} < \min(0.01, 0.1 \cdot d_{intra})$, where the contact density is defined as the fraction of residue pairs of native distance below 8Å. At the final stage, the database includes 750 Pfam domain families matching 13,156 PDB with a total number of 77,109 intra-chain structure units and 54,065 inter-chain structure units.

As we will see in the *Results* section, too small interfaces are not detectable by DCA. Finally, we have 750 Pfam families, which form our data set.

Besides these homo-oligomeric families, we collect also a database of all those families, which fulfill our constraints on the sequence number and the existence of a high-resolution PDB structure, but which are not classified as homo-oligomeric biological assembly or do not pass the tests on the interface size. These are a total of 1234 families.

## *Direct-Coupling Analysis*

Direct-Coupling Analysis is based on the maximum-entropy modeling of protein families, with the aim to reproduce amino-acid frequencies in single MSA columns, and frequencies of amino-acid co-occurrences in pairs of MSA columns, via an otherwise unconstrained statistical model. This leads, for an aligned amino-acid sequence $(A_1, ..., A_L)$ of length L to a generalized Potts model, or Markov Random Field,

$$P(A_1,...,A_L) = \frac{1}{Z} \exp \left\{ \sum_{i<j} e_{ij}(A_i, A_j) + \sum_i h_i(A_i) \right\}.$$

Parameters of this model, independently for each Pfam family, are calculated using the pseudo-likelihood maximization approach of (31) with standard settings for the reweighting and the regularization parameters. Following the same reference, coupling strengths are characterized by the Frobenius norm $F_{ij} = \sum_{a,b} [e_{ij}(a,b)]^2$ with subsequent average-product correction $F_{APC,ij} = F_{ij} - F_{\bullet j} F_{i \bullet} / F_{\bullet \bullet}$. The dot indicates an average over the concerned position(s). These $F_{APC}$ are used to sort all residue pairs, strong DCA couplings are interpreted as predictors for residue-residue contacts inside or between proteins.

## *Structure Prediction*

We used HADDOCK (37) (high ambiguity driven protein-protein docking) to obtain the complex/dimers of RR_REC domains. HADDOCK is based on "sticky" surface regions between the provided chains.

Briefly, we isolated chain A from a representative PDB file for the different response regulators Transreg_c (PDB:1nxs), LytTR (PDB:4cbv) and GerE (PDB:4e7p). The resulting monomers were docked in HADDOCK by using all amino acids involved in predicted contacts (Table 1) as surface residues. This results in several clusters of docked structures for the three response regulators (Transreg_c: 2, GerE: 2, LytTR: 3), which differ substantially. By counting the number of correctly realized DCA-contacts for the different clusters, one can in each case identify a unique conformation in best agreement with the DCA-contact predictions.

We stress that the used prediction technique differs from prior docking studies using coevolutionary direct information (20). Here, we use the DCA-contact predictions not as N pairwise constraints/ energetic biases between residues (i,j) in the initial docking step but as one large surficial cluster formed by all involved residues in any predicted contact (i,j). In a second step we discriminate between the predicted clusters by counting the number of specific contacts (I,j) as selection criteria. The main advantage of this two-step approach is to search a wider range of conformations in the initial step, before deciding on the specific conformation in best agreement with the DCA-contacts. We also performed docking studies, which used the DCA-contacts as direct constraints with $_E$SBMTools (39) more similar to prior work (cf. *Supplement*).

## ACKNOWLEDGEMENTS

GU and MW acknowledge funding by the Agence Nationale de la Recherche via the project COEVSTAT (ANR-13-BS04-0012-01). SJL and AS recognize from the Impuls- und Vernetzungsfond of the Helmholtz association. AS recognizes support from a Google Faculty Research Award. HS was funded by grant GM106085 from the National Institute of General Medical Sciences, National Institutes of Health, USA.

**Figures**

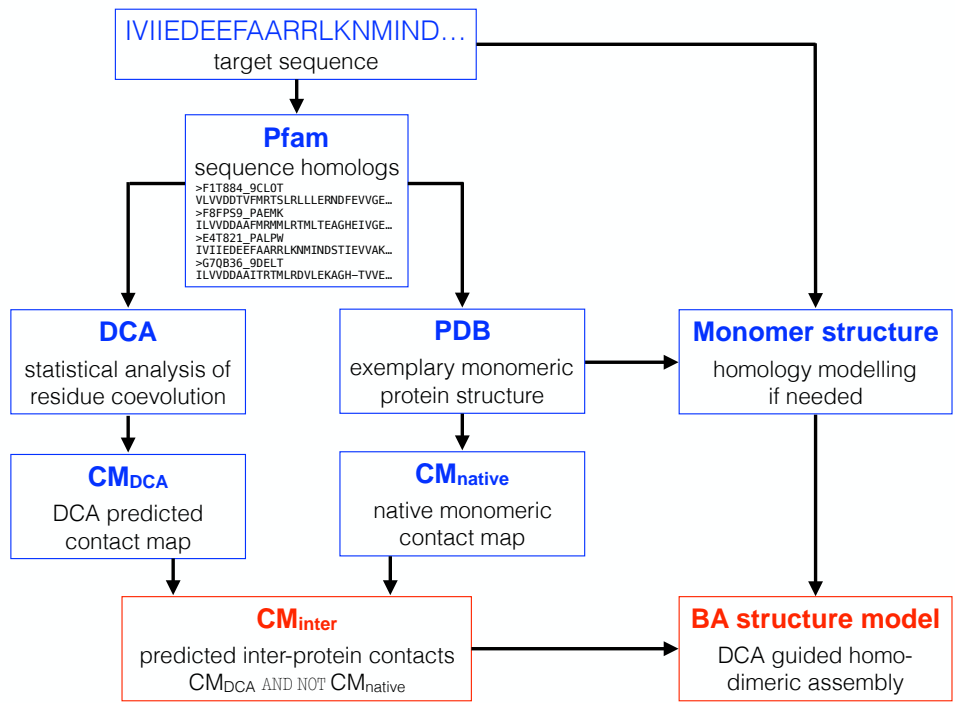

**Figure 1. Analysis approach** – Starting from a target amino-acid sequence, we use PFAM to extract a multiple-sequence alignment (MSA) of homologous sequences, and exemplary monomer protein structures from the PDB. DCA is run on the MSA to predict a contact map. Inter-protein residue contacts are obtained by pruning all those DCA-predicted residue pairs, which are compatible with the known monomeric structure. To actually obtain a structural model for the biological assembly, we use two monomers (possibly homology modeled if not directly available in the PDB) and dock them guided by the DCA inter-protein predictions.

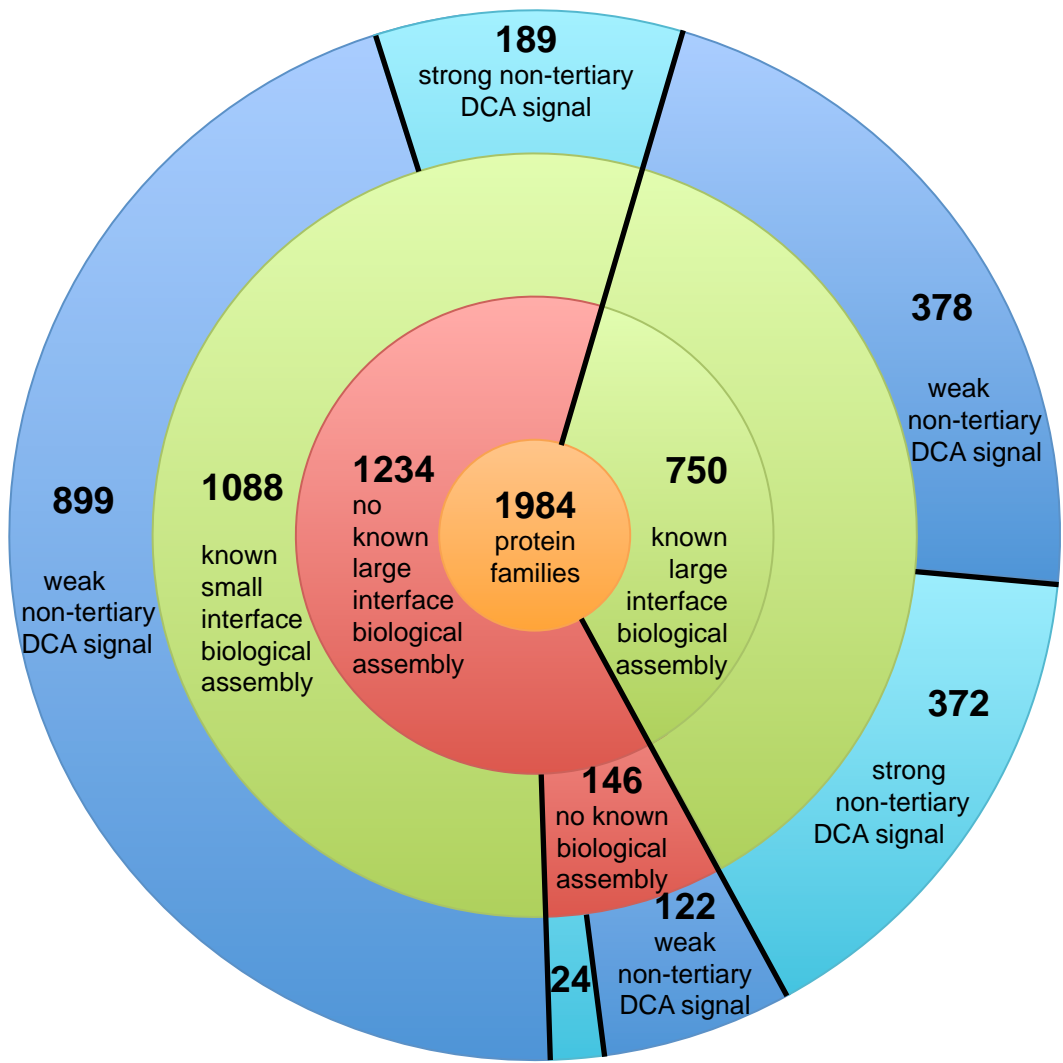

**Figure 2. Overview of results for investigated protein families –** To understand the effect of residue coevolution in homo-oligomerization, we investigate in total 1,984 protein families (PF) from the PFAM database. They fulfill the requirements of sufficient sequence information and a known monomeric structure, which is a requirement to differentiate coevolving intra-protein-contacts (i.e. tertiary contacts) from putative inter-protein-contacts (i.e. quaternary contacts). Larger assemblies have stronger signals, as only some residues forming the interface strongly co-evolve. We thus subdivide the PF into two classes pending on their known interface sizes. 750 PF have a large known interface in their biological assembly. For 372 PF of these, we can identify this interface through strongly coevolving residue pairs purely based on sequences. For the other 1,234 PF, no large biomolecular assemblies are known. 1,088 PF, however, form a known small interface, and this can be identified for 189 PF. For the 146 PF possessing no known interface, the large majority does not show strongly coevolving residue pairs. For 24, however, we find strongly co-evolving pairs indicating a possible biological homo-oligomeric assembly.

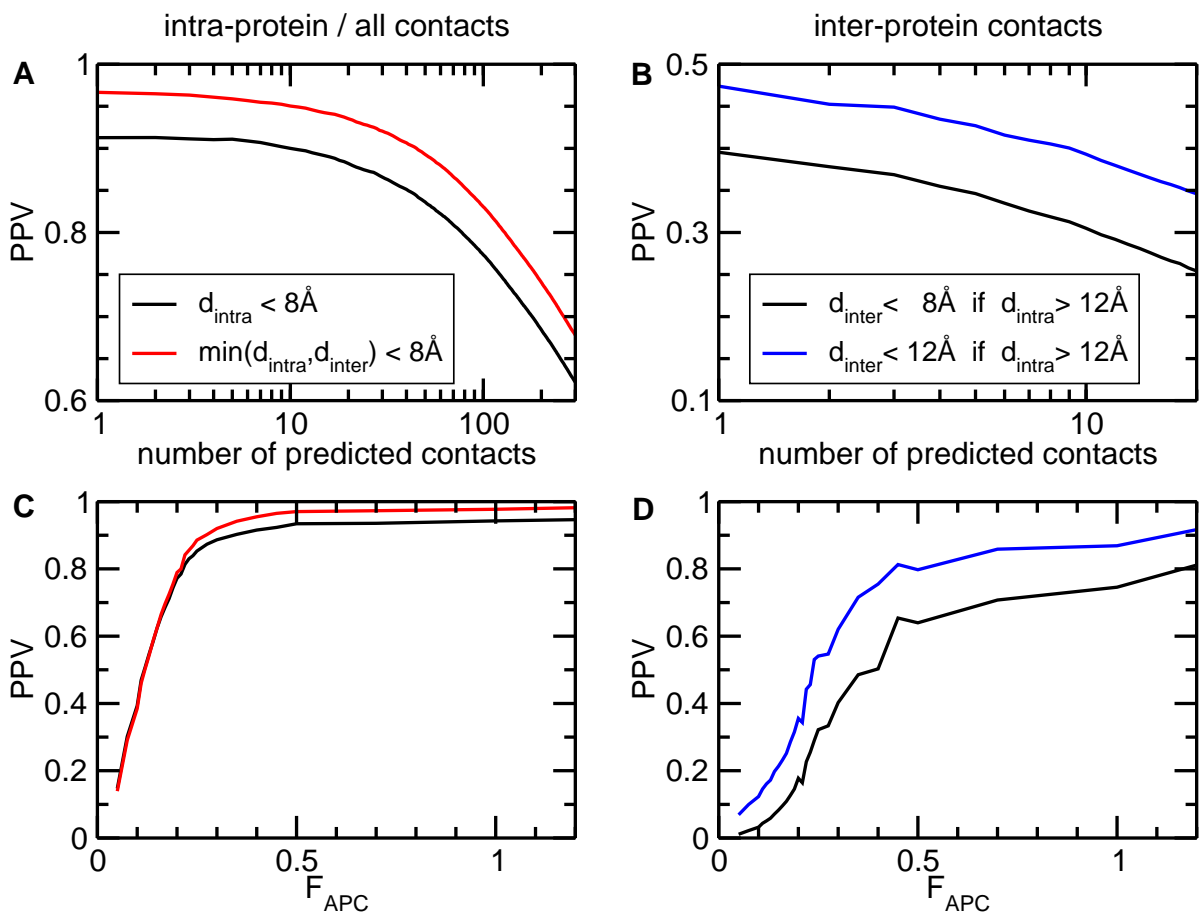

**Figure 3. Average prediction accuracy for intra- and inter-protein contacts** – The figures report the performance of DCA for predicting intra- and inter-protein contacts for the 750 selected families forming homo-oligomers. Panel A shows the positive predictive value (PPV = TP/(TP+FP) = number of true postive predictions / number of predictions) in dependance of the number of prediction, averaged over all 750 families. The black line takes into account contacts inside the monomer, whereas the red line counts also intra-protein contacts extracted from biological assemblies deposited in the PDB. While already the intra-protein contacts prediction is of high accuracy (average PPV 90.2% for first 10 predictions per family, 77.5% for first 100 predictions), we find that a large fraction of "false positives" are actually inter-protein contacts (combined PPV 95.2% for 10 predictions, 83.2% for 100 predictions per family). Panel C traces (same colours as in Panel A) the PPV as a function of the DCA-score $F_{APC}$, it shows that almost constantly high PPV above 90% are obtained for $F_{APC}$ above 0.3. Panel C and D analyze only predictions, which are highly incompatible with the monomeric structure (minimum atom distance above 12Å). We see that almost 40% are in contact between the proteins (8Å distance cutoff), and more than 46% are in spatial vicinity (12Å cutoff). We find again that high-scoring residue pairs with $F_{APC}>0.3$ have a high probability to be in inter-protein contact (average PPV above 80%).

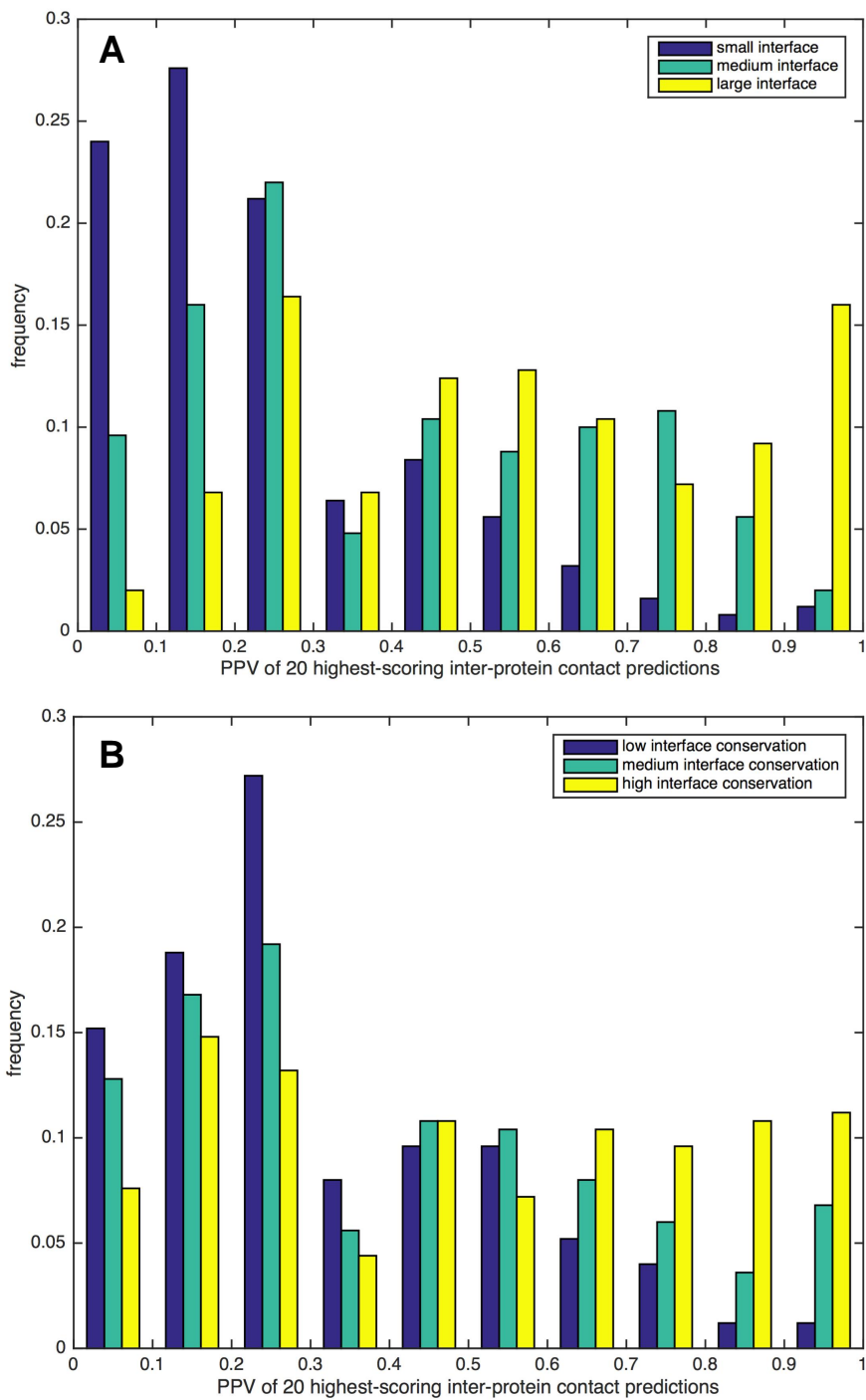

**Figure 4. Influence of interface size and conservation on the accuracy of inter-protein contact prediction** – Panel A shows a histogram for the PPV (=TP/(TP+FP)) of the first 20 contact predictions for the 750 selected oligomer-forming protein families. For the histogram, the dataset is divided into three classes according to their interface size (small, medium, large), measured by the number of inter-protein contacts. While families with large interface size dominate the highest-accuracy cases, families with small interfaces typically lead to very bad positive predictive values. Panel B shows an analogous histogram, now with the dataset of selected families with at least two alternative homo-oligomeric structure. More conserved interfaces (i.e. with less variability between the different representative structures of the concerned family) frequently show higher PPV values than variable interfaces. Since the coevolutionary signal for each oligomerization mode is specific to an appropriate selection of proteins, and become weak in the overall family alignment, which mixes different oligomerization modes.

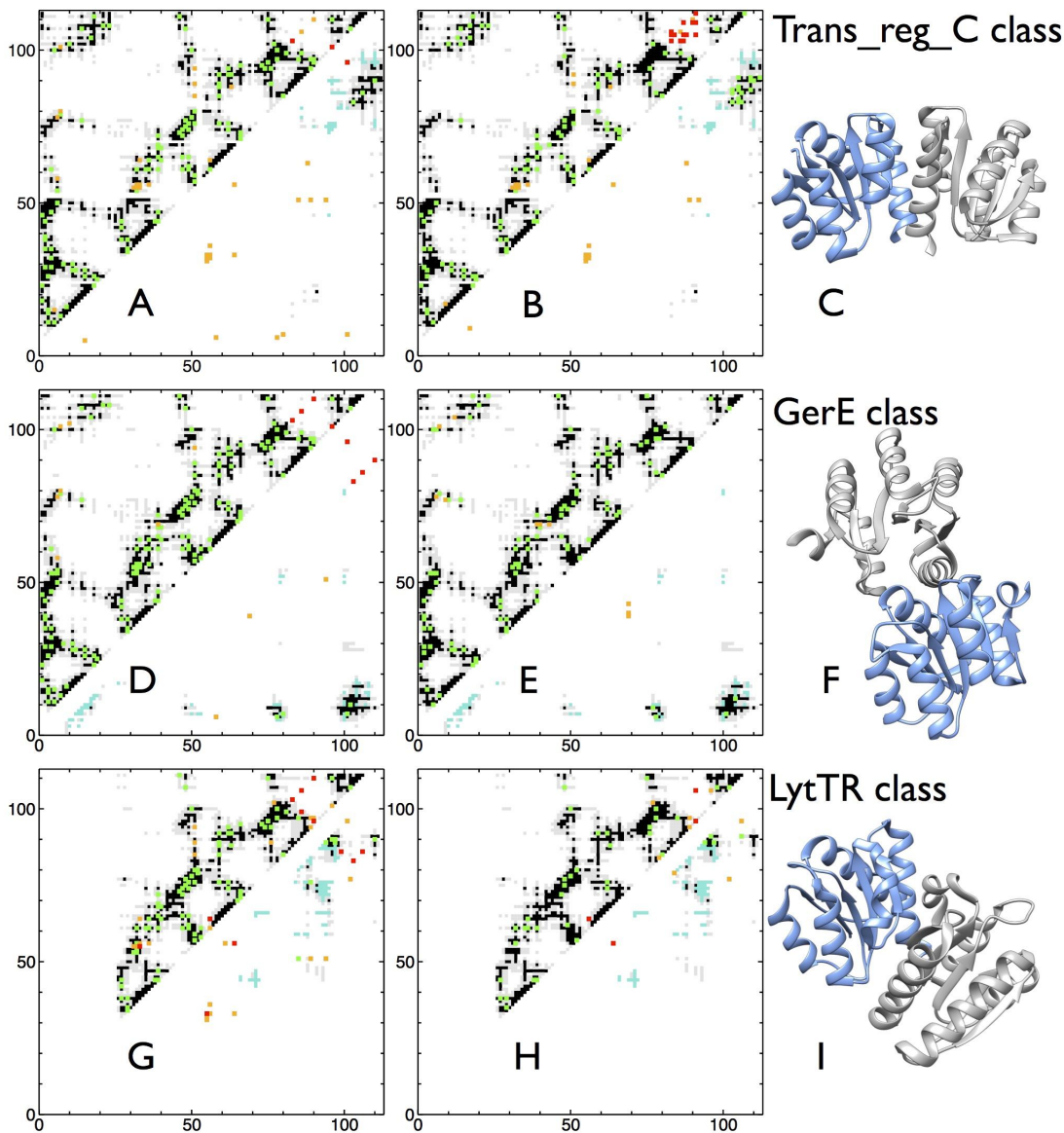

**Figure 5. Intra- and inter-protein contacts prediction for different classes of response regulator** – Response regulators (RR) with DNA-binding domains dimerize when activated, but the dimerization mode differs between different DNA-binding domain families. Here we analyse RR having the three major DNA-binding domains, Trans_Reg_C, GerE or LytTR. As the right panels (C,F,I) show, they have substantially differing biological assemblies, the blue domain is positioned such that a visual comparison is facilitated. The left panels (A,D,G) show the contact prediction using the full RR Pfam family (PF00072), the middle panels (B,E,H) the prediction using only the sub-MSA of RR containing the corresponding DNA-binding domain. The upper-diagonal triangles show the intra-protein contacts, native contacts (d<8Å) are depicted by black dots, residue pairs with native distances below 12A by grey dots. The colored dots correspond to DCA-predictions with $F_{APC}>0.3$, green dots are within 8A distance, orange dots between 8 and 12Å, red dots above 12Å. The lower diagonal shows the inter-protein contact map. Black / grey dots correspond to distances below 8Å / 12Å, which are not in contact inside the monomer, and thus potentially detectable by our method. Light-blue dots correspond to inter-protein distances below 12Å which are not detectable by our method since they are contained also in the intra-protein contact map. The other colored dots report all DCA predictions with $F_{APC}>0.3$ and $d_{intra} > 8A$: green if $d_{inter}<12Å$, orange if $d_{inter}>12Å$ but $d_{intra}<12Å$, and red if $\min(d_{inter},d_{intra})>12Å$. While the full Pfam alignment results only in very few strong coevolutionary signals with incompatibility to the monomeric structure (one residue pair with $F_{APC}>0.3$ and $d_{intra}>8Å$ in all three structures), the sub-class alignment result in strongly improved inter-protein contact predictions. Green, orange and red dots ($F_{APC}>0.3$ and $d_{intra}>8A$) were used in the docking simulations.

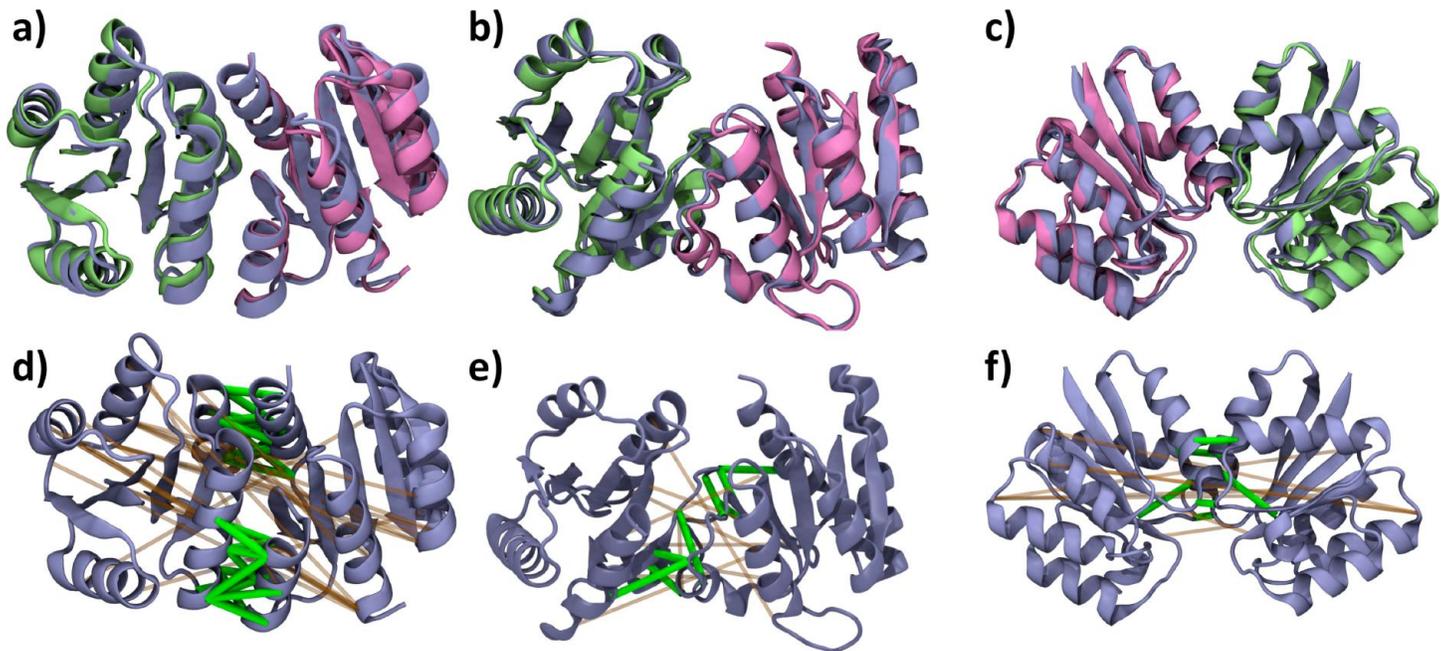

**Figure 6. Prediction of homodimeric structures** – The predicted interdomain contacts (cf. Fig. 5) allow reliable predictions of homodimeric structures with up to crystal structure accuracy for the different response regulators OmpR (PDB 1nxs; a,d), LytTR (PDB 4cbv; b,e) and GerE (PDB 4e7p; c,f). Displayed are the predicted homodimers (blue) overlaid with their known crystal structures (a,b,c); monomers highlighted as green and purple in a, b, c). The Backbone-Rmsd's between the prediction and the crystal structures are 1.1Å (OmpR), 1.1Å (GerE) and 1.2Å (LyTR). Interestingly, for LyTR we predict the biological assembly 2. In the predictions, only part of the predicted interdomain contacts are realized ($d_{inter} < 8$Å; highlighted green in d, e, f) while many predicted contacts are not formed (orange in d, e, f). All these contacts, however, are close to being realized in the monomer as intra contacts within a threshold of 12Å ($d_{inter}>8$Å; $d_{intra}<12$Å). None of the predicted contacts are > 12Å both inter and intra.

**Table 1:** Investigated Response Regulators, their predicted DCA-contacts and the contacts realized in the structure prediction of their homo-dimeric form

| Protein | Predicted Contacts ($F_{apc}>0.3$; $d_{intra}>8$A; ordered by DCA-score) | Realized Contacts in Homodimer ($d_{inter}<8$A) |
|---|---|---|
| OmpR (PDB 1nxs) | 84-104, 87-107, 33-57, 34-58, 35-57, 58-66, 84-106, 38-58, 88-106, 85-106, 92-113, 91-110, 91-111, 92-106, 87-104, 35-58, 92-110, 88-104, 65-89, 88-110, 89-106, 84-107, 34-57, 53-90, 53-93, 12-20 | 84-104, 87-107, 84-106, 88-106, 85-106, 92-113, 91-110, 91-111, 92-106, 87-104, 92-110, 88-104, 88-110, 84-107 |
| LyTR (PDB 4cbv) | 101-120, 87-112, 66-74, 89-94, 99-104, 101-106, 83-87, 100-104, 99-107 | 101-120, 101-106, 99-107 |
| GerE (PDB 4e7p) | 11-80, 42-72, 14-107, 43-72, 8-81, 46-72, 9-104 | 11-80, 14-107, 9-104 |